\documentclass{aa}

\usepackage{graphicx}
\usepackage{natbib}
\usepackage{txfonts}
\bibpunct{(}{)}{;}{a}{}{,}

\begin{document}

  \title{Center-limb dependence of photospheric velocities in regions of emerging magnetic fields on the Sun}

   \author{A. Khlystova
           \inst{1}
          }
   \institute{Institute of Solar-Terrestrial Physics, Russia, 664033, Irkutsk p/o box 291; Lermontov st., 126a\\
              \email{hlystova@iszf.irk.ru}
             }

  \date{Received ; accepted }

\abstract
{}
{We investigate the ratio of the vertical and horizontal velocities of the photospheric plasma flows in the region of emerging magnetic fields on the Sun.}
{We carried out a study of photospheric velocities during the first hours of the appearance of 83 active regions with magnetic flux more than 10$^{21}$ Mx with data acquired by the Michelson Doppler Imager (MDI) on board the Solar and Heliospheric Observatory (SOHO). The emerging magnetic fluxes under investigation were isolated from extended concentrations of existing magnetic fields; they have different spatial scales and are located at different distances from the solar disk center.}
{We found that the values of maximum negative Doppler velocities that accompany the emergence of active region magnetic fields during the first 12 hours increase nonlinearly with the heliocentric angle. This result shows that the horizontal photospheric velocities of plasma outflows are higher than the vertical ones of the plasma upflows during the first hours of the emergence of active regions. The horizontal velocity component at the site of emerging active regions exceeds that of convective flows in the quiet Sun. A comparison between the velocities and the maximum value of the total magnetic flux has not revealed any relation.}
{}

\maketitle

\section{Introduction}

The emerging magnetic flux of active regions on the solar surface appears in the form
of separate loops; they then coalesce to form opposite polarity poles that
separate and increase in size \citep{str99}. Earlier studies of velocities of
plasma motions were carried out in the region of emerging magnetic fields at
different height levels of the Sun. Investigations into the chromospheric level
have been made by \citet{bru67,bru69} and others \citep[see review
by][]{cho93}. Velocities at the photospheric level were measured with different
methods (discussed in more detail below). In the last decade, helioseismological
methods have been applied to study the subphotospheric level
\citep{cha99,kos00,zha08,kos09,kom08,kom09}.

Direct measurements of photospheric velocities revealed negative values (blue
Doppler shift) on the polarity inversion line of emerging magnetic fields. The
events under consideration were located in the central part of the solar disk,
and, therefore, vertical motions were measured. Negative velocities or plasma
upflows up to 1 km s$^{-1}$ were observed in separate magnetic loops inside
emerging active regions \citep{tar90,lit98,str99,kub03,gri09}.
\citet{gri07} revealed high velocities of about 1.7 km s$^{-1}$ at the
beginning of the powerful active region emergence NOAA 10488 at heliographic
coordinates N08 E31 ($B_{0}$ = + 4.9).

Indirect measurements of the horizontal velocities of the photospheric flow that accompanies
the emergence of active regions were made by tracking the displacement of individual
magnetic elements. The velocities obtained ranged from 0.1 to 1.4 km s$^{-1}$
\citep{fra72,sch73,str99}. \citet{bar90} studied 45 bipolar pairs in
the emerging active region. The opposite polarity poles were separating from
each other with velocities of 0.5--3.5 km s$^{-1}$, decreasing with time
(consequently, the drift with regard to the polarity inversion
line is only half that value). \citet{gri09} calculated the separation velocities of the external
boundaries of the photospheric magnetic flux in the active region NOAA 10488.
The velocities decreased as the magnetic fields emerged: they were 2--2.5 km\,s$^{-1}$ 
by the end of the first hour and 0.3 km s$^{-1}$ in two hours and a
half.

It should be noted that there are several papers dealing with the study of
vertical \citep{gug06} and horizontal \citep{har73,cho87,hag01}
velocities at  photospheric level during the appearance of ephemeral active
regions. We do not discuss these works in detail, since they concern another
spatial scale of emerging magnetic fields.

The papers above are the researches into photospheric velocities that accompany the emergence of magnetic fields in active regions located only in the central zone of the solar disk. Plasma motions in the emerging magnetic fields distant from the disk center have never been studied before.

\section{Data processing and investigated objects}

\begin{figure*}
\centering
\includegraphics[width=\textwidth]{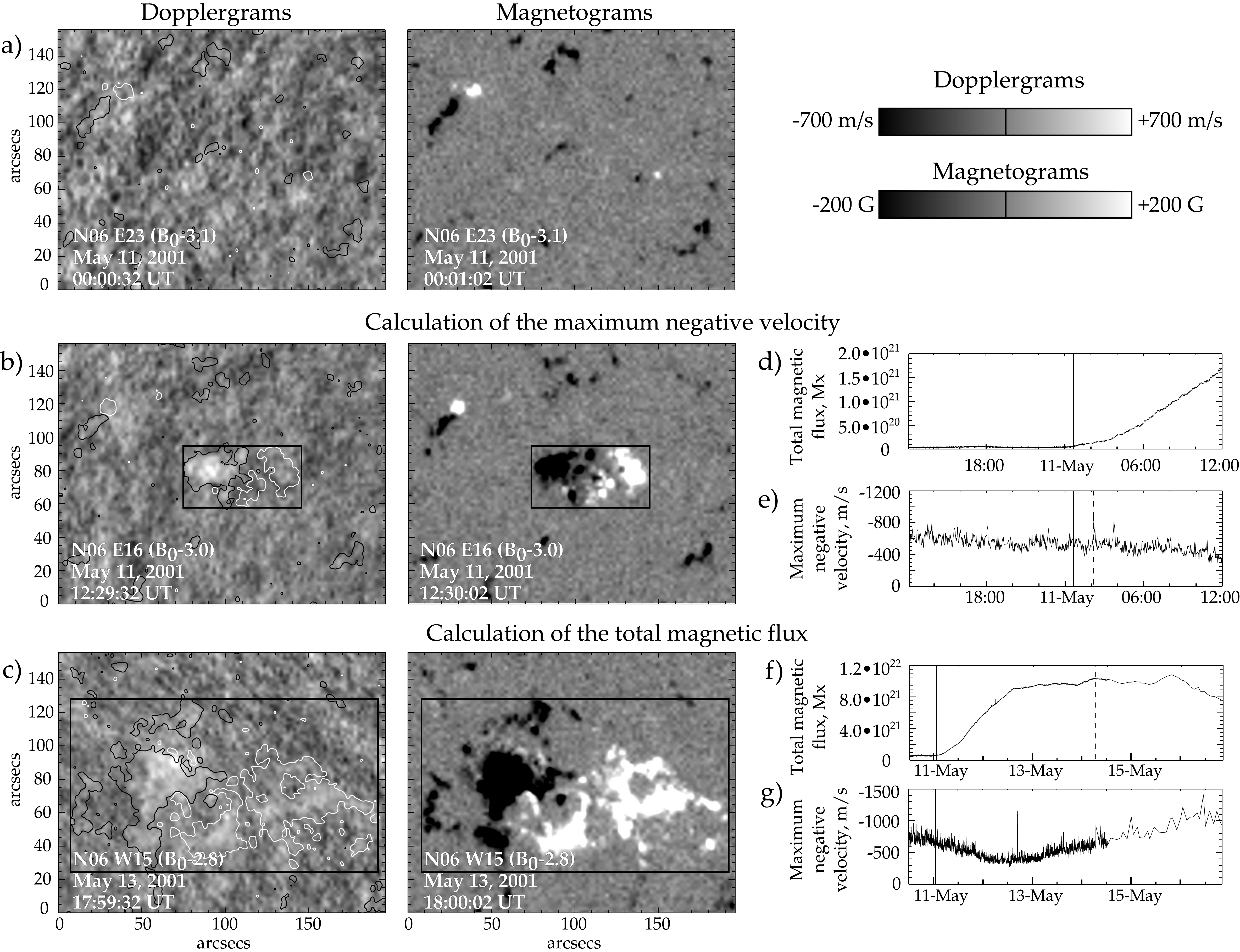}
\caption{Active region NOAA 9456: (a) the solar surface 30 minutes before the magnetic field emergence; (b) 12 hours after the beginning of the emergence; (c) near the maximum of evolution. The left column represents Dopplergrams with magnetic field isolines +60 G (white line) and -60 G (black line); the positive velocities correspond to the motion of matter away from the observer, while the negative velocities correspond to the motion of matter toward the observer. The right column represents magnetograms of the longitudinal field; the black color corresponds to the negative magnetic field, while the white one corresponds to the positive magnetic field. The small rectangle marks the calculation area of the maximum negative velocity during the first 12 hours of the magnetic field emergence; the big one indicates the calculation area of the total magnetic flux. Plots: (d), (e) variation in the parameters investigated during the first 12 hours of the magnetic field emergence; (f), (g) variation in the parameters investigated until the maximum of the active region evolution. The beginning of the active region appearance is marked by the vertical solid line; the extreme values of the maximum negative velocity in plot (e) and the total magnetic flux in plot (f) are shown by the vertical dashed line.}
\label{fig1}
\end{figure*}

We used full solar disk magnetograms and Dopplergrams in the photospheric line Ni I 6768
$\mbox{\AA}$ and continuum images obtained on board the space observatory
SOHO/MDI \citep{sch95}. The temporal resolution of the magnetograms
and Dopplergrams is 1 minute, that of the continuum, 96 minutes. The spatial
resolution of the data is 4$\arcsec$, the pixel size is approximately
2$\arcsec$. Magnetograms with a 1.8.2 calibration level were used
\citep{ulr09}. Besides photospheric velocities of solar plasma motions, the
Dopplergrams include the contribution of 1) the velocity of the differential solar
rotation; 2) the velocity of the SOHO satellite with regard to the Sun; 3) the instrumental
distortions caused by nonuniform transmission of MDI's filter systems on the
field of view. The technique described in \citet{gri07} was used to separate
photospheric velocities. The negative velocity on SOHO/MDI Dopplergrams
corresponds to the blue Doppler shift (motion of matter toward the observer);
the positive one, to the red Doppler shift (motion of matter away from the
observer). There are regular errors in velocity measurements at different distances from the disk center. For example, Doppler velocities near the limb are underestimated by up to 40 m s$^{-1}$ because of limb darkening \citep{alb85}. Additionally, we deal with different height levels above the solar surface in measurements of velocities at the disk center and near the limb because of the different optical thickness of the atmosphere.

We study the velocities that accompany the emerging magnetic fields.
Therefore, the precise spatial superimposition of the used data is vary
important. For this purpose, the region of the emerging magnetic fields was taken
from the time sequences of magnetograms and Dopplergrams, taking into account
solar rotation. An approximate value of the displacement of the region was
calculated with the differential rotation law for photospheric magnetic fields
\citep{sno83}. The exact tracking of the region under investigation was
performed by applying two magnetograms adjacent in time, with the use of
cross--correlation analysis. This procedure requires the existence of the magnetic poles which slightly vary in time on the magnetograms. The heliographic coordinates obtained were used to crop identical fragments with a size of
160$\arcsec\times$160$\arcsec$ (or 80$\times$80 pixels) from the magnetogram and
Dopplergram obtained at the same time. Thus, we achieved the precise spatial
superimposition of the data. It has allowed us to draw reliable
conclusions about the processes that take place. For the active region emerging
near the limb, we selected the cropped region in a way that it excludes the area outside the limb.

Time variations of the total magnetic flux of the active region and the maximum negative
velocities were calculated with the use of a sequence of the magnetogram and
Dopplergram fragments corresponding to the emerging active region. The
calculation area was limited to the region of the emerging magnetic fields (Fig.
~\ref{fig1}). The position of the limits was visually controlled.

The plasma motions connected with the emerging magnetic fields are observed on the
background of the convective flows. If the average velocity or the velocity flux
is considered as a characteristic of motions of matter, the center--limb
dependence of convective velocities will make a substantial contribution.
Besides, the active regions under consideration have different spatial scales
and are located at different distances from the solar disk center; therefore
the solar surface regions where the calculation is performed have different sizes.
This complicates a comparison of the emerging active regions that are located at different distances from the
disk center. The maximum negative velocity (motion of matter toward the observer) is taken as a characteristic of plasma
motion in the regions of the emerging magnetic fields during the first 12 hours
(Fig. ~\ref{fig1}b). The magnetic field emergence begins with the appearance of the loop
apex where the magnetic field is horizontal; therefore, we began the calculation of the maximum
velocity values 30 minutes before the beginning of the appearance of
active region. The magnetic flux of the active region
emerges as separate loops, and the maximum negative velocity corresponds to the
appearance of a single magnetic loop. Therefore, the chosen parameter does not
characterize the emerging active region as a whole, it only shows the highest
velocity. For example, two peaks corresponding to the emergence of two different
magnetic loops can be distinguished in the plot of variations in the maximum
negative velocity during the first hours of the NOAA\,9456 appearance (Fig. ~\ref{fig1}e).
The residual magnetic flux emerges at lower velocities. To exclude Evershed flows we also took care of the time of the
penumbra formation in sunspots, using  continuum images. The Evershed flows are horizontal photospheric outflows in a sunspot penumbra. Their maximum velocity values can reach 2 km s$^{-1}$ in SOHO/MDI low spatial resolution data \citep{bai98}.

The total magnetic flux is calculated inside isolines $\pm$60 G taking into account the
projective effect (Fig. ~\ref{fig1}c)

\begin{eqnarray}
\Phi & = & |\sum_{i=1}^{n_{+}}(B_{i} \cdot S_{i})| + |\sum_{i=1}^{n_{-}}(B_{i}
\cdot S_{i})|,
\end{eqnarray}
where $\Phi$ is the total magnetic flux in Mx, $B_{i}$ is the magnetic field
induction in G, $S_{i}$ is the area on the solar surface corresponding to $i-th$
pixel, $n_{+}$ is the number of pixels inside isoline +60\,G (positive polarity)
and $n_{-}$ is the number of pixels inside isoline -60\,G (negative polarity).
The isoline level of 60\,G was chosen to exclude both the magnetogram noise (about 30\,G)
and the contribution of the magnetic field with a short lifetime.
Sometimes small magnetic loops appear in the place of the future active region
several hours prior to the main emergence of the magnetic fields. The beginning
of the active region emergence was taken as a time moment corresponding to the
beginning of the continuous growth of magnetic flux. We determined the maximum value of the total
magnetic flux by the first inflection point of the curve where
the flux increase was followed by its decrease (Fig. ~\ref{fig1}f) or by the last value of
the plot (for active regions passing beyond the west--limb or with
insufficiently downloaded data sequences). Note that powerful active regions
significantly increase in size as they develop, and the extended concentrations of the
magnetic fields existing on the surface enter into the region of new flux
emergence (Fig. ~\ref{fig1}a, ~\ref{fig1}b, and ~\ref{fig1}c). Interaction between existing and emerging
magnetic fields is accompanied by integration or cancellation processes and,
hence, by increase or decrease in the magnetic flux. In order to take this into account to some extent, the signal background that existed before the emergence of magnetic fields was subtracted from the total magnetic flux maximum in all active regions.

\begin{figure}
\centering
\includegraphics[width=8.86cm]{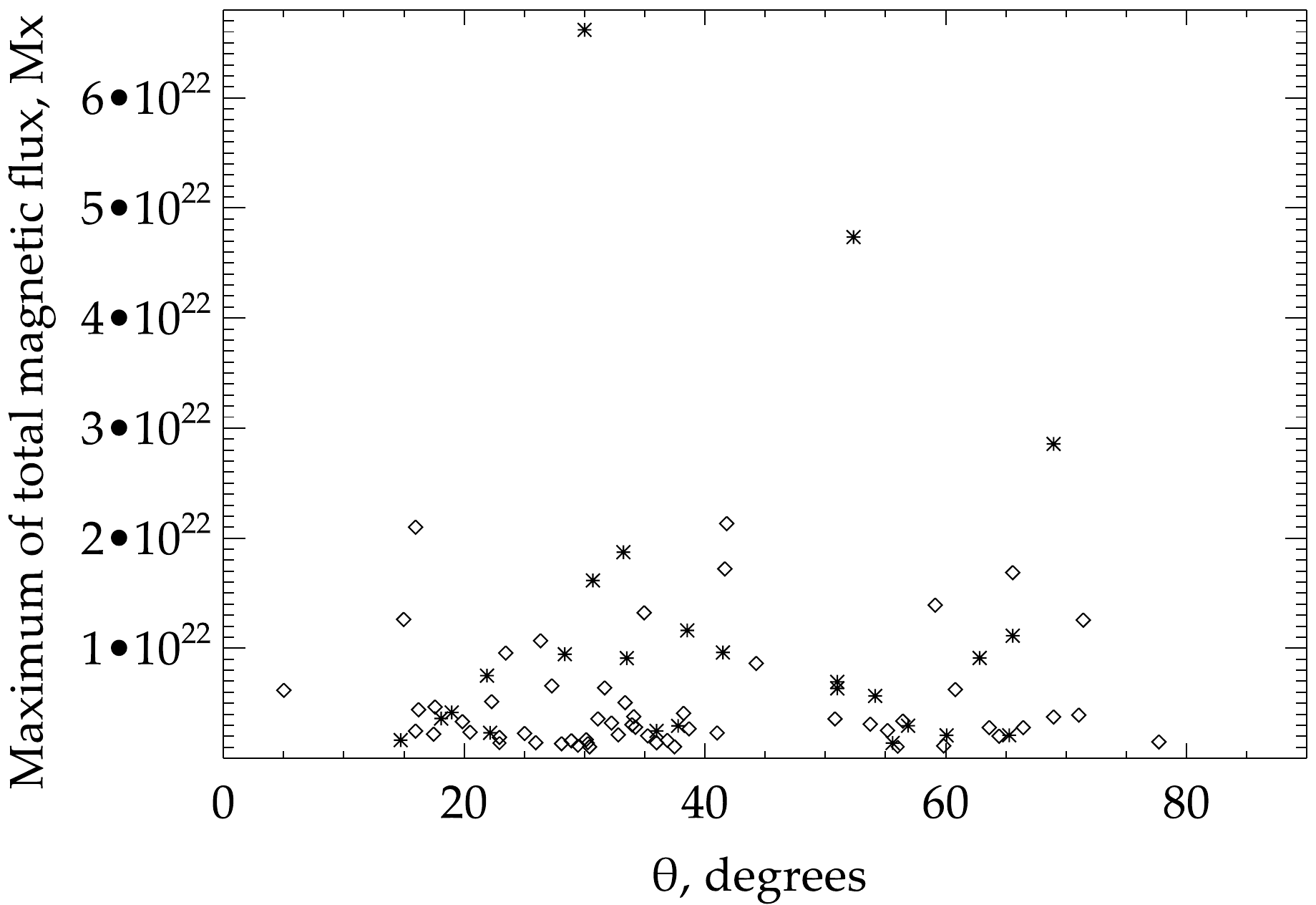}
\caption{Total magnetic flux of the active regions under investigation
versus the heliocentric angle that corresponds to the beginning of the magnetic
field emergence. The total magnetic flux maximum was determined by the
inflection point in the increase of the flux curve (marked by diamonds) or by
the last measurement (marked by asterisks). Active regions with magnetic saturation are also marked by asterisks. In active regions marked by asterisks, the maximum of total magnetic flux may be higher.}
\label{fig2}
\end{figure}

\begin{figure*}
\centering
\includegraphics[width=\textwidth]{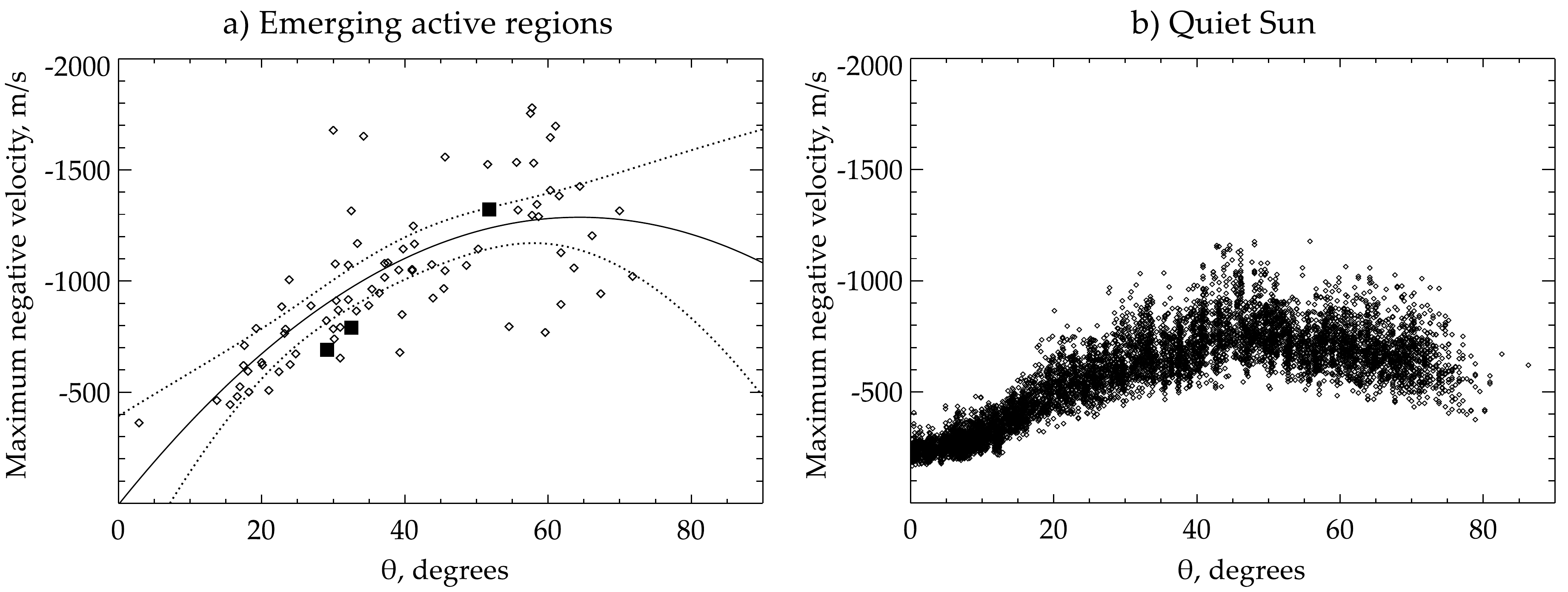}
\caption{(a) Maximum negative velocities during the first 12 hours of the emergence of active regions versus the heliocentric angle that corresponds to the position of the active regions at this point in time. Diamonds mark normal active regions; black squares mark inverse ones. The solid line corresponds to the nonlinear regress equation (2); dotted lines correspond to 99$\%$ confidence intervals for the means. (b) Maximum negative velocities of convective flows in the quiet Sun versus the heliocentric angle.}
\label{fig3}
\end{figure*}

The selection of active regions for this investigation was based on the following criteria. Active regions have to emerge on the visible side of the solar disk. Emerging magnetic fields have to be isolated from the extended concentrations of existing magnetic fields. The presence of single poles with a total magnetic flux of less than 0.5$\times$10$^{21}$ Mx was allowed in the area of the direct emergence of an active region during the first 12 hours. The objects under investigation have to have complete data series with a one--minute time resolution during the first 12 hours of appearance. With this in mind, the 83 active regions were selected for the period of 1999--2008. Among them, there were 80 active regions with a normal magnetic field configuration and three with an inverse magnetic field configuration according to Hale's polarity law. The events had different spatial scales and were located at different distances from the solar disk center (Fig. ~\ref{fig2}). The total magnetic flux of each event exceeded 10$^{21}$ Mx. Magnetic saturation is possible in SOHO/MDI measurements of the magnetic field strength. This occurs when the spectral line is shifted out of the filter system passband. \citet{liu07} modeled MDI measurements and demonstrated that magnetic saturation takes place only in strong magnetic fields with high velocities. According to their calculations, magnetic saturation is reached at the magnetic field strength of 2800 G, when the velocity is $\pm$2000 m s$^{-1}$. In our study, the maximum magnetic field strength exceeds 2800 G only in two events of the active regions under consideration. These events are also marked by asterisks in Fig. ~\ref{fig2}. The morphology of the active regions (i.e., the degree of fragmentation of magnetic fields, the rate and monotony of magnetic flux increase, and the orientation of the axis connecting opposite polarity poles with regard to the line of sight) was not taken into account.

\section{Results}

Figure ~\ref{fig3}a presents the maximum negative velocities observed during the first 12 hours emergence of active regions according to their position with regard to the disk center. The distance from the disk center is expressed in the heliocentric angle $\theta$ -- it is the angle between the normal to the surface and the line--of--sight to the emerging magnetic flux. The maximum negative velocities obtained are observed, as a rule, on the polarity inversion line of the emerging magnetic fields; however, they are related to the convection in adjacent regions for some events in the central part of the disk ($\theta<25^\circ$). Figure ~\ref{fig3}a shows that values of maximum negative velocities increase nonlinearly with the heliocentric angle. The mean velocity for $\theta=20^\circ$ is -673 m s$^{-1}$; the deviation from the mean does not exceed 200 m s$^{-1}$. The mean velocity for $\theta=60^\circ$ is -1277 m s$^{-1}$; the deviation from the mean reaches up to 550 m s$^{-1}$. One can see a growth tendency of deviation from mean with increasing $\theta$ (Fig. ~\ref{fig3}a). The high deviation for large $\theta$ indicates the existence of other relations that determe the value of the horizontal velocity of plasma flow at the site of emerging magnetic fields at the photospheric level. Our statistics include three inverse active regions whose parameters do not deviate from the dependence formed by normal active regions. Thus we see that the horizontal velocities of plasma outflows exceed the vertical ones of plasma upflows at the beginning of the emergence of active regions.

Figure ~\ref{fig3}b shows the maximum negative velocities for the quiet Sun connected with convective flows. To calculate this dependence, the region on the surface was traced taking into account solar rotation. The region with a size of 40$\arcsec\times40\arcsec$ under study was going through the longitude range of W00--W67 at the latitude of S04 from 3 to 8 May 1999. The tilt of the solar North rotational axis toward the observer $B_{0}$ was -3.5$^\circ$. Data with a one-minute resolution were used; they were processed with the technique described in section 2 of this paper. In Fig. ~\ref{fig3}b one can see that the velocities increase with the heliocentric angle; these values do not exceed 1200 m s$^{-1}$. The comparison between the plots in Fig. ~\ref{fig3}a and Fig. ~\ref{fig3}b has revealed that the horizontal velocity component at the sites of the emergence of active regions is higher than that of the convective flow in the quiet Sun.

We performed the regression analysis with the second degree polynom. As a result, we obtained the equation

\begin{eqnarray}
\upsilon & = & 6.01 - 40.18 \theta + 0.31 \theta^2,
\end{eqnarray}
where $\upsilon$ is the maximum negative Doppler velocity during the first 12 hours of the emergence of active regions in m\,s$^{-1}$ and $\theta$ is the heliocentric angle in degrees. According to the F--statistic this regression model is significant. The correlation ratio (estimate of the closeness of the nonlinear relation) between the value of the maximum negative velocity and the heliocentric angle $\theta$ (or the angle of view to the emerging magnetic flux) is equal to -0.74; this implies a high relation between the parameters under consideration. The confidence intervals for the means were calculated with a confidence probability of 99$\%$. They are marked by dotted lines in Fig. ~\ref{fig3}a. The confidence intervals have close limits for $10^\circ<\theta<70^\circ$ with high statistics and show that equation (2) approximates the data well. The limits of the confidence intervals essentially increase for $\theta<10^\circ$ and $70^\circ<\theta$ with low statistics. It follows from equation (2) that the mean vertical velocity for $\theta=0^\circ$ is 6 m s$^{-1}$ with a confidence interval of $\pm$399 m\,s$^{-1}$. Positive velocity values are false in this confidence range. Obviously, velocities corresponding to $\theta=0^\circ$ at the sites of emerging active regions will be not lower than the velocities of convective flows of the quiet Sun ($\sim$-150 m s$^{-1}$). The absence of statistics for $\theta<10^\circ$ at the emerging active regions does not allow one to take it into account in regression equation (2). The mean horizontal velocity for $\theta=90^\circ$ from equation (2) is -1081.44 m s$^{-1}$ with a confidence interval $\pm$601 m s$^{-1}$. The wide confidential interval shows that this value of the mean horizontal velocity is ambiguous.

\begin{figure}
\centering
\includegraphics[width=8.86cm]{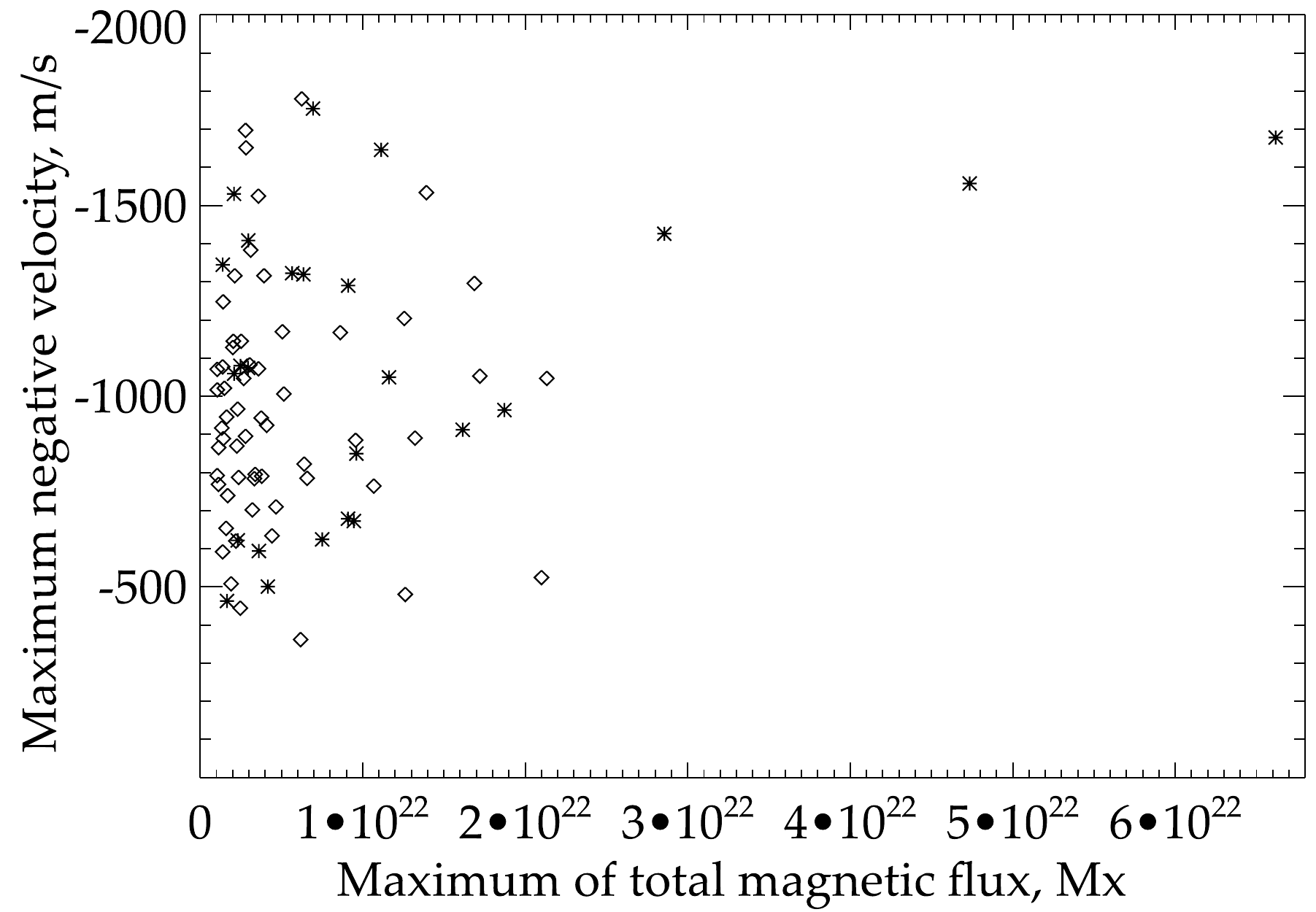}
\caption{Maximum negative velocities during the first 12 hours of the magnetic
field emergence versus the maximum value of the total magnetic flux in the
active regions. The designations are the same as for the Fig. ~\ref{fig2}.}
\label{fig4}
\end{figure}

A comparison of the maximum negative velocities with the total magnetic flux did not reveal any relation (Fig. ~\ref{fig4}). The comparison between the plots in Fig. ~\ref{fig2} and Fig. ~\ref{fig3}a also showed no connection, because high velocities were not observed in emerging active regions with high magnetic flux in the central part of the solar disk, whereas near the limb there may be high velocity values even in active regions with low magnetic flux. From the author's point of view the absence of a dependence between velocity and total magnetic flux can be explained by the fact that active region magnetic fields emerge by separate fragments on time scales from several hours to 5-7 days and therefore the maximum negative velocity during the first 12 hours does not characterize an active region emergence as a whole.

\section{Conclusions}

We presented a statistical study of the velocities of plasma flows that accompany the emergence of active regions during the first 12 hours at the photospheric level with data with high temporal resolution. We found that the values of maximum negative Doppler velocities increase nonlinearly with the heliocentric angle. This shows that the horizontal velocities of plasma outflows exceed the vertical ones of plasma upflows during the first hours of active region emergence. Horizontal velocities at the sites of emerging active regions are higher than those of convective flows in the quiet Sun. This result is a direct confirmation of theoretical models that showed that magnetic fields emerging in the solar atmosphere expand faster in the horizontal direction than in the vertical direction \citep[e.g.][]{shi89,fan01,mag03,arc04}.

\begin{acknowledgements}
The author is grateful to the referee for helpful comments that improved the manuscript.
This work used data obtained by the SOHO/MDI
instrument. SOHO is a mission of international cooperation between ESA and NASA.
The Michelson Doppler Imager is a project of the Stanford--Lockheed Institute
for Space Research. I am grateful to my supervisors V.M. Grigor'ev and L.V.
Ermakova for their helpful suggestions on this scientific research, and to V.G.
Fainshtein for very useful discussions. I would like to thank Tom Duvall and
Alexander Kosovichev for discussions that helped me to understand peculiarities of
SOHO/MDI Dopplergrams better. This study was supported by RFBR grants
08-02-00027-a, 09-02-00165-a, 10-02-00607-a, 10-02-00960-a, the NNSFC-RFBR grant
08-02-92211, state contract 02.740.11.0576 and by the Russian Academy of Sciences
(the program 4 of the RAS Presidium and the joint project 4 of its Siberian Branch).
\end{acknowledgements}

\bibliographystyle{aa}
\bibliography{ref}

\end{document}